\documentclass[preprint,12pt]{elsarticle}

\usepackage{amssymb}
\usepackage{amsmath}
\usepackage{longtable}
\usepackage{lineno,hyperref}
\usepackage[flushleft]{threeparttable}
\usepackage{multirow}
\usepackage{lscape}
\usepackage{xcolor}
\definecolor{blue}{rgb}{0,0,1}

\usepackage{tikz}
\usetikzlibrary{positioning}
\usetikzlibrary{shapes.geometric, arrows}

\journal{International Journal of Electrical Power \& Energy Systems}

\begin{document}
\bibliographystyle{elsarticle-num}
\begin{frontmatter}

\title{Tight MILP Formulation for Pipeline Gas Flow with Linepack}

 \author[1,2]{T. Klatzer\corref{correspondingauthor}}
 \cortext[correspondingauthor]{Corresponding author:}
 \ead[url]{thomas.klatzer@tugraz.at}
 \author[1,2]{S. Wogrin}
 \author[3,4]{D. A. Tejada-Arango}
 \author[3,5]{G. Morales-Espa{$\tilde{\textrm{n}}$}a}
 
 \affiliation[1]{organization={Institute of Electricity Economics and Energy Innovation (IEE), Graz University of Technology},
            addressline={Inffeldgasse 18}, 
            city={Graz},
            postcode={8010},
            country={Austria}}

 \affiliation[2]{organization={Research Center ENERGETIC, Graz University of Technology},
             addressline={Rechbauerstraße 12},
             city={Graz},
             postcode={8010},
             country={Austria}}

 \affiliation[3]{organization={TNO Energy \& Materials Transition},
             addressline={Radarweg 60},
             city={Amsterdam},
             postcode={1043 NT},
             country={The Netherlands}}

 \affiliation[4]{organization={Instituto de Investigación Tecnológica, Escuela Técnica Superior de Ingeniería, Universidad Pontificia Comillas},
             addressline={C. Alberto Aguilera 23},
             city={Madrid},
             postcode={28015},
             country={Spain}}
             
 \affiliation[5]{organization={Faculty of Electrical Engineering, Mathematics and Computer Science, Delft University of Technology},
             addressline={Mekelweg 5},
             city={Delft},
             postcode={2628 CD},
             country={The Netherlands}}

%% Abstract
\begin{abstract}
In integrated power and gas energy system optimization models (ESOMs), pipeline gas transmission with linepack is a particularly complex problem due to its non-linear and non-convex character. For ESOMs based on mixed-integer linear programing, piecewise linearization is a well-established convexification approach for this problem, which, however, requires binary variables to model feasible combinations of linear gas flow and pressure segments and thus can quickly become computationally challenging. 
In order to improve computational performance, this paper proposes a piecewise linearization method specifically designed to be tight, resulting in a reduced problem space a solver can explore faster.
We provide numerical results comparing the proposed formulation against two piecewise linearizations from the literature, both from a theoretical point of view and in terms of practical computational performance, with results showing an average speed-up of 2.57 times for our case study.
Test cases are carried out on a modified 24-bus IEEE Reliability Test System and a 12-node gas system, considering discrete unit commitment decisions.
\end{abstract}

%% Keywords
\begin{keyword}
Integrated energy system modeling \sep gas flow \sep linepack \sep piecewise linearization \sep MILP
\end{keyword}

\end{frontmatter}



\section*{Nomenclature}
\label{sec:Nomenclature}
Sets
\begin{longtable}[l]{ l l }
$\mathcal{K}$       & Set of hours $k$.                                                         \\
$\mathcal{G}$       & Set of all generators $g$.                                                \\
$\mathcal{T}$       & Set of gas-fired generators $t$.                                          \\
$\mathcal{R}$       & Set of renewable generators $r$.                                          \\
$\mathcal{I}$       & Set of power buses $i$.                                                   \\
$\mathcal{S}$       & Set of gas sources $s$.                                                   \\
$\mathcal{M}$       & Set of gas nodes $m,n$.                                                   \\
$\mathcal{L}$       & Set of pipelines $l$ connecting node $m$ and $n$.                         \\
$\mathcal{C}$       & Set of compressors $c$ connecting node $m$ and $n$.                       \\
$\mathcal{A}^{\mathcal{T}}_m$   & Set of gas-fired generators connected to $m$.                 \\
$\mathcal{A}^{\mathcal{S}}_m$   & Set of gas sources connected to $m$.                          \\
$\mathcal{Z}$       & Set of grid points $z,\Tilde{z}$ for piecewise linearization.             \\
$\mathcal{U}$       & Set for piecewise linearization with elements $u$, $v$.                   \\  
$\mathcal{W}$       & Set for piecewise linearization with elements $w$.                   
\end{longtable}
\addtocounter{table}{-1}

Parameters
\begin{longtable}[l]{ l l }
$C^{G}_{s}$                             & Supply cost of gas source $s$ (\$/MSm\textsuperscript{3}).                            \\
$C^{OM}_{g}$                            & O\&M cost of generator $g$ (\$/MWh).                                                  \\
$C^{EN,GN}$                             & Cost of electricity, gas non-supplied (\$/MWh), (\$/MSm\textsuperscript{3}).          \\
$D^{E}_{k,i}$                           & Electricity demand at time $k$ and bus $i$ (MW).                                      \\
$D^{G}_{k,m}$                           & Gas demand at time $k$ and node $m$ (MSm\textsuperscript{3}/h).                       \\
$\overline{P}^{G}_{g}$                  & Max. production of gas source $s$ (MSm\textsuperscript{3}/h).                         \\
$\underline{P}_{g},\overline{P}_{g}$    & Min./Max. output power of generator $g$ (MW).                                         \\
$RU_t,RD_t$                             & Ramp-up/down rate of generator $t$ (MW).                                              \\
$CS^{V}_{t}$                            & Variable gas consumption of generator $t$ (p.u.).                                     \\
$H^{G}$                                 & Heating value of gas (MWh/MSm\textsuperscript{3}).                                    \\
$R^{G}_{l}$                             & Pipeline parameter of $l$ ((MSm\textsuperscript{3}/h barg)\textsuperscript{2}).       \\
$LP_{l}$                                & Linepack factor of pipeline $l$ (MSm\textsuperscript{3}/barg).                        \\
$LP^{ini}_{l}$                          & Initial linepack of pipeline $l$ (MSm\textsuperscript{3}/barg).                       \\
$\eta_l$                                & Efficiency factor for gas loss of pipeline $l$ (p.u.).                                \\
$\Lambda_c$                             & Compression ratio of compressor $c$ (p.u.).                                           \\
$\overline{F}^{C}_{c}$                  & Max. compressor capacity of compressor $c$ (MSm\textsuperscript{3}/h).                \\
$CS^{G}_{c}$                            & Gas consumption of compressor $c$ (p.u.).                                             \\
$F_{l,z},P_{l,z}$                       & Flow, pressure of pipeline $l$ at grid point $z$ (MSm\textsuperscript{3}/h), (barg).  \\
$P^{\Delta}_{l,z}$                      & Pressure difference of pipeline $l$ at grid point~$z$ (barg).                         \\
$\underline{P}_{m},\overline{P}_{m}$    & Min./Max. gas pressure at node $m$ (barg).                                    
\end{longtable}
\addtocounter{table}{-1}

Variables
\begin{longtable}[l]{ l l }
$p^{G}_{k,s}$               & Gas production at time~$k$ of gas source~$s$ (MSm\textsuperscript{3}/h).                  \\  
$p^{E}_{k,g}$               & Power generation at time~$k$ of generator~$g$ (MW).                                       \\
$\hat{p}_{k,t}$             & Power output at time~$k$ above technical minimum of~$t$ (MW).                             \\
$cs^{G}_{k,t}$              & Gas consumption at time~$k$ of generator~$t$ (MSm\textsuperscript{3}/h).                  \\
$y_{k,t}$                   & Startup at time~$k$ of generator~$t$, $\in\{0,1\}$.                                       \\
$u_{k,t}$                   & Commitment at time~$k$ of generator~$t$, $\in\{0,1\}$.                                    \\
$z_{k,t}$                   & Shutdown at time~$k$ of generator~$t$, $\in\{0,1\}$.                                      \\
$p_{k,m}$                   & Gas pressure at time~$k$ at node~$m$ (barg).                                              \\
$p^{+}_{k,l},p^{-}_{k,l}$   & Forward/reverse pressure difference at time~$k$ of pipeline~$l$ (barg).                   \\
$lp_{k,l}$                  & Linepack at time~$k$ of pipeline~$l$ (MSm\textsuperscript{3}).                            \\
$f_{k,l}$                   & Average gas flow at time~$k$ of pipeline~$l$ (MSm\textsuperscript{3}/h).                  \\
$f^{+}_{k,l},f^{-}_{k,l}$   & Forward/reverse average gas flow at time~$k$ of~$l$ (MSm\textsuperscript{3}/h).           \\
$f^{In,Out}_{k,l}$          & Gas inflow and outflow at time~$k$ of pipeline~$l$ (MSm\textsuperscript{3}/h).            \\
$f^{C}_{k,c}$               & Gas flow at time~$k$ of compressor~$c$ (MSm\textsuperscript{3}/h).                        \\
$\gamma_{k,l,z}$            & Filling at time~$k$ of pipeline~$l$ of grid point~$z$ (p.u.).                             \\
$\delta_{k,l,z}$            & Forces filling at time~$k$ of pipeline~$l$ of adjacent~$z$, $\in\{0,1\}$.                 \\
$\xi_{k,l}$                 & Gas flow direction at time~$k$ for pipeline~$l$, $\in\{0,1\}$.                            \\
$ns^E_{k,i}$                & Electricity non-supplied at time~$k$ and bus~$i$ (MW).                                    \\
$ns^G_{k,m}$                & Gas non-supplied at time~$k$ and node~$m$ (MSm\textsuperscript{3}/h).                                 
\end{longtable}
\addtocounter{table}{-1}

\section{Introduction}
\label{sec:Intro}
\subsection{Motivation and Literature Review}
\label{sec:Motivation&LitRev}
Decarbonization of energy systems is one of the pressing challenges of the twenty-first century.
A highly renewable power sector and the electrification of energy demands are at the heart of this transformation. However, for applications and processes where electrification is not possible or efficient, e.g., for hard-to-abate industries or storing large quantities of energy for long duration~\cite{EUH2Strategy2020}, gases like hydrogen, ammonia, synthetic natural gas etc. will be key. 
As the production of large quantities of these gases from renewable power and their bulk transmission are envisioned -- with projected investment costs of several hundred billion, e.g.,~\cite{EUH2Strategy2020} -- co-optimization of power and gas systems is becoming more important~\cite{Fodstad2022,ILM2024}.

In this context, energy system optimization models (ESOMs) can be valuable planning tools.
ESOMs typically face the challenge of striking a balance between the representation of the technical, temporal and spatial domains, while providing globally optimal solutions and remaining computationally tractable~\cite{Kotzur2021}.
In integrated power and gas ESOMs, pipeline gas transmission is a particularly complex problem in the technical domain due to the nonlinear and nonconvex relation of (bidirectional) gas flows and gas pressure~\cite{Correa-Posada2015}. In addition, the slow gas flow dynamics, i.e., the time difference before a change in outflowing gas mass is reflected at the pipeline inlet, enables pipelines to be used for short-term gas storage (linepack). Linepack is particularly important to mitigate operational uncertainty, e.g., for balancing weather-induced gas demand variations~\cite{Kazda2020}, and as a source of flexibility for the power system, e.g., for frequently dispatched gas-fired power plants~\cite{Raheli2024}, or for buffering (offshore) hydrogen production in the future~\cite{Bødal2024}.

In the literature, there is a plethora of modeling approaches for the pipeline gas transmission problem with linepack. Typically, the methods with the highest level of physical detail, e.g., directly solving the problem via nonlinear programming~\cite{Gao2024}, sequential linear programming~\cite{Qadrdan2014,Liu2019a}, mixed-integer conic relaxation~\cite{Borraz-Sanchez2016,Schwele2019} etc., cannot guarantee finding globally optimal solutions and/or are computationally expensive. In contrast, relaxation-based methods, such as polyhedral envelopes~\cite{Mhanna2022}, show relatively fast computational performance but can significantly violate physical realities, i.e., gas flowing against the pressure difference. Raheli et al.~\cite{Raheli2024} provides a detailed overview of these and other methods.

Under the paradigms of global optimality and physical feasibility, outer approximation based on Taylor series and piecewise linearization methods have been suggested. Commonly, these methods require formulation as a mixed-integer linear program (MILP). For the outer approximation based on the Taylor series~\cite{He2018a,Ordoudis2019,Shin2022}, binary variables are required to model bidirectional gas flows using big-Ms, while piecewise linearization methods, which have been reviewed in~\cite{Vielma2010} and applied in the context of oil production~\cite{Silva2014} and hydropower scheduling~\cite{Brito2020}, typically use binary or SOS2-type variables to select linear segments to approximate the average gas flow along a pipeline and pressures at its start and end points~\cite{Correa-Posada2015,Correa-Posada2014,Shao2017}.

From a computational point of view, solving MILPs, which are generally NP-hard problems~\cite{Conforti2014}, has been improved immensely over the past 20 years, with an average total speed-up of 1,000 due to combined improvement of solvers and hardware~\cite{Koch2022}. Nevertheless, how a MILP is formulated can still significantly improve (or harm) its solution time~\cite{Hooker2011}. Typically, the solution time of an MILP is improved the closer (tighter) its relaxed solution is to its MILP solution~\cite{Morales2013}. This is because the solver has to explore less space between the relaxed feasible solution and the integer feasible solution~\cite{Morales2013}. In addition, tightness not only helps to obtain higher quality MILP solutions, i.e., a reduced optimality gap, in a given time but also higher quality relaxed solutions since relaxed integer variables take values closer to their integer values.
The second important characteristic alongside tightness is compactness, which corresponds to the number of constraints and nonzeros in the MILP formulation. In general, a MILP can be tightened by adding strong valid inequalities, i.e., cuts. Adding constraints, however, implies repeatedly solving larger relaxed LP instances during the branch and bound procedure, so compromising compactness can also result in longer solution times compared to less tight MILP formulations. In a nutshell: Tightening typically harms compactness, while a more compact formulation is typically less tight~\cite{Morales2013}. Ultimately, the highest quality MILP formulations are both tight \textit{and} compact simultaneously. 

Considering the above, MILP formulations using outer approximation based on Taylor series are not tight, since they use big-Ms, which are not tight by definition, i.e., the purpose of big-Ms is to ensure that specific constraints are non-binding.
In contrast,  piecewise linearization methods studied in~\cite{Correa-Posada2014}, e.g., incremental (INC) or SOS2, are tight  formulations (convex hulls), which benefits their computational performance. The crux is that this is only valid for the linearization, i.e., INC or SOS2, per se. The union of two convex hulls, however, do not result in convex hulls. For the pipeline gas transmission problem with linepack, which requires linearization of average gas flow \textit{and} nodal gas pressures, this implies that the computational benefit due to the tight character of the linearization is expected to be diminished.

\subsection{Contributions}
\label{Contributions}
In this paper, we propose a piecewise linearization method (Z) that is tailored to the pipeline gas transmission problem with linepack and specifically designed to be tight. Compared to the INC and SOS2 linearizations, Z is a more conservative approximation of the general gas flow equation, particularly at high pressures. Therefore, the feasible regions of INC/SOS2 and Z are not identical, i.e., this can result in a moderate difference in gas flows and/or pressures. However, our numerical results show that the computational speed-up under Z is significant compared to INC or SOS2.

\noindent The contributions of this paper are:
\begin{itemize}
    \item We propose the Z linearization method based on the forward (reverse) average gas flow and pressure difference along a pipeline. For symmetrical bidirectional gas flows, this results in a more compact formulation regarding the number of variables compared to INC and SOS2\footnote{This is generally the case, except for some special network configurations, i.e., number of gas nodes $\ll$ number of pipelines (i.e., many parallel pipelines connecting the same gas nodes) while the number of linear segments is also very small.}.
    \item We show numerically that Z is tighter than INC and SOS2 by computing the fractional vertices of their relaxed LP feasible region.
    \item We illustrate the computational speed-up of Z over INC and SOS2 numerically in a number of case studies of an integrated power and gas ESOM. 
\end{itemize}

\subsection{Paper Organization}
\label{Organization}
The remainder of this paper is structured as follows: In Section \ref{sec:Model} we describe the integrated power and gas ESOM; Section \ref{sec:Linearization} presents the concept and mathematical framework of the proposed Z linearization; and Section \ref{sec:Numerical} compares numerical results of Z, SOS2 and INC with respect to tightness, problem size (compactness), computational performance and operational results. Finally, Section~\ref{sec:Conclusions} concludes the paper.

\section{Integrated Power and Gas Optimization Model}
\label{sec:Model}
This section introduces the mathematical formulation of the integrated power and gas optimization model. Since this paper focuses on gas flow modeling with linepack, constraints solely relevant to the power sector, e.g., generation from renewables, the power balance etc., are omitted here. For a detailed overview of these constraints, the interested reader is referred to state-of-the-art power system models, e.g.,~\cite{Wogrin2022-LEGO}.
\subsection{Objective Function and Bounds}
\label{sec:OF}
The optimization model aims to minimize the operational cost of power and gas supply.
The objective function~(\ref{eqn:OF}) includes:
(\textit{i})    the cost of supplying gas to the system;
(\textit{ii})   operation and maintenance (OM) costs of gas-fired power plants;
(\textit{iii})  OM costs of renewable generation units;
(\textit{iv})   cost of non-supplied electricity; and 
(\textit{v})    cost of non-supplied gas.
Constraints \eqref{eqn:PNS}-\eqref{eqn:GNS} limit the amount of non-supplied electricity and gas per power bus and gas node respectively.
\begin{align}
     % objective function (OF) ---------------------------------------------
    min \sum_{k\in\mathcal{K}} \Big(  \sum_{s\in\mathcal{S}} \underbrace{  C^{G}_s  p^{G}_{k,s}}_{(\textit{i})}
                                          +\sum_{t\in\mathcal{T}} \underbrace{  C^{OM}_t p^{E}_{k,t}}_{(\textit{ii})} 
                                          +\sum_{r\in\mathcal{R}} \underbrace{  C^{OM}_r p^{E}_{k,r}}_{(\textit{iii})} \nonumber \\ 
                                          +\sum_{i\in\mathcal{I}} \underbrace{  C^{EN}  ns^{E}_{k,i}}_{(\textit{iv})}      
                                          +\sum_{m\in\mathcal{M}} \underbrace{  C^{GN}  ns^{G}_{k,m}}_{(\textit{v})} \Big)
\label{eqn:OF}
\end{align}
%other constraints
\begin{align}
   0 \leq ns^{E}_{k,i} \leq D^E_{k,i} \quad \forall~k\in\mathcal{K},~i\in\mathcal{I} \label{eqn:PNS}    
\end{align}
\begin{align}
   0 \leq ns^{G}_{k,m} \leq D^G_{k,m} \quad \forall~k\in\mathcal{K},~m\in\mathcal{M} \label{eqn:GNS}
\end{align}

\subsection{Gas Production and Consumption}
\label{sec:Operation}
The total gas demand of the integrated power and gas system is supplied by gas sources \eqref{eqn:Gas-Sources}, which, for the sake of simplicity, are only constrained by an upper production limit.
\begin{align}
  0 \leq p^{G}_{k,s} \leq \overline{P}^{G}_{s} \quad \forall~k\in\mathcal{K},~s\in\mathcal{S} \label{eqn:Gas-Sources}
\end{align}
Besides the gas demand for heating and industrial processes $D^G_{k,m}$, gas-fired power plants consume gas to produce power. 
Gas-fired units are modeled by a standard unit commitment (UC) problem \cite{Wogrin2020} where: the total power generation $p^E_{k,t}$ is defined as the sum of the technical minimum and the power output above the technical minimum $\hat{p}_{k,t}$~\eqref{eqn:UCTotOut}; the power output above the technical minimum is zero in case of startup~\eqref{eqn:UCMaxOut1} and shutdown decisions~\eqref{eqn:UCMaxOut2}; ramp-up~\eqref{eqn:UCRampUp} and ramp-down rates \eqref{eqn:UCRampDw} are limited; and the commitment, startup and shutdown decisions are related~\eqref{eqn:UCLogic}-\eqref{eqn:UCLogicDW}; and defined as binary variables~\eqref{eqn:UCBinary}. Finally, in addition to a standard UC problem in the power sector,~\eqref{eqn:UCConversionP} establishes a relation between the total power generation and the gas consumption of the unit.
%--------------------------------------------------------------------------
%               Gas thermals
\begin{align}
    p^E_{k,t} =  u_{k,t} \underline{P}_t + \hat{p}_{k,t} 
    \quad \forall~k\in\mathcal{K},~t\in\mathcal{T}
    \label{eqn:UCTotOut}
\end{align}
\begin{align}
    \hat{p}_{k,t} \leq (\overline{P}_t - \underline{P}_t)(u_{k,t} - y_{k,t})
    \quad \forall~k\in\mathcal{K},~t\in\mathcal{T}
    \label{eqn:UCMaxOut1}
\end{align}
\begin{align}
    \hat{p}_{k,t} \leq (\overline{P}_t - \underline{P}_t)(u_{k,t} - z_{k+1,t})
    \quad \forall~k\in\mathcal{K},~t\in\mathcal{T}
    \label{eqn:UCMaxOut2}
\end{align}
\begin{align}
    \hat{p}_{k,t} - \hat{p}_{k-1,t} \leq u_{k,t} RU_t
    \quad \forall~k\in\mathcal{K},~t\in\mathcal{T}
    \label{eqn:UCRampUp}
\end{align}
\begin{align}
    \hat{p}_{k,t} - \hat{p}_{k-1,t} \geq -u_{k-1,t} RD_t
    \quad \forall~k\in\mathcal{K},~t\in\mathcal{T}
    \label{eqn:UCRampDw}
\end{align}
\begin{align}
    u_{k,t} - u_{k-1,t} = y_{k,t} - z_{k,t}
    \quad \forall~k\in\mathcal{K},~t\in\mathcal{T}
    \label{eqn:UCLogic}
\end{align}
\begin{align}
    y_{k,t} \leq u_{k,t}
    \quad \forall~k\in\mathcal{K},~t\in\mathcal{T}
    \label{eqn:UCLogicUP}
\end{align}
\begin{align}
    z_{k,t} \leq 1-u_{k,t}
    \quad \forall~k\in\mathcal{K},~t\in\mathcal{T}
    \label{eqn:UCLogicDW}
\end{align}
\begin{align}
    u_{k,t}, y_{k,t}, z_{k,t} \in \{0,1\} 
    \quad \forall~k\in\mathcal{K},~t\in\mathcal{T}
    \label{eqn:UCBinary}
\end{align}
\begin{align}
    cs^{G}_{k,t} H^{G}  = p^{E}_{k,t} CS^{V}_t
    \quad \forall~k\in\mathcal{K},~t\in\mathcal{T}
    \label{eqn:UCConversionP}
\end{align}
\subsection{Gas Transmission}
\label{sec:GasTrans}
The gas flow in high-pressure transmission pipelines follows a set of partial differential equations~(PDEs).
In the context of optimization models, the discretized PDEs (dynamic gas flow model) are commonly reduced to the quasi-dynamic gas flow model by neglecting the impacts of inertia and kinetic energy on the gas flow~\cite{Correa-Posada2015}. In the quasi-dynamic model,~\eqref{eqn:GasFlow} relates the (bidirectional) squared average gas flow $f_{k,l}|f_{k,l}|$ (where $|\cdot|$ is the absolute function) and the squared nodal pressures $p^2_{k,m}$ and $p^2_{k,n}$ via the constant $R^{G}_{l}$, which comprises pipeline and gas characteristics~\cite{Correa-Posada2015}. Note that in the ESOM,~\eqref{eqn:GasFlow} is modeled as~\eqref{eqn:EQ3}-\eqref{eqn:IEQ6}.
\begin{align}
    % relation gas flow and pressure
    f_{k,l} |f_{k,l}| = R^{G}_{l}(p^{2}_{k,m} - p^{2}_{k,n})
    \quad \forall~k\in\mathcal{K},~l(m,n)\in\mathcal{L}
    \label{eqn:GasFlow}
\end{align}
The amount of gas stored in a pipeline is referred to as linepack.
Although the mathematical framework for including linepack in optimization models has been described in previous works, e.g.,~\cite{Correa-Posada2015}, below we provide a brief overview for context.

Linepack $lp_{k,l}$ is modeled as the product of the average pressure along the pipeline and a linepack factor $LP_{l}$~\eqref{eqn:Linepack}.
\begin{align}
    lp_{k,l} = LP_{l} \frac{p_{k,m} + p_{k,n}}{2}
    \quad \forall~k\in\mathcal{K},~l(m,n)\in\mathcal{L}
    \label{eqn:Linepack}
\end{align}
It follows a state of charge concept, which relates the linepack of two consecutive time steps (including gas losses represented by the efficiency $\eta_l$), the gas inflow $f^{In}_{k,l}$ and the outflow $f^{Out}_{k,l}$~\eqref{eqn:LinepackSoC}. Constraint~\eqref{eqn:LinepackFinSoC} enforces equality of initial and final linepack.
\begin{align}
    lp_{k,l} = lp_{k-1,l} \eta_{l} + f^{In}_{k,l} - f^{Out}_{k,l}
    \quad \forall~k\in\mathcal{K},~l(m,n)\in\mathcal{L}
    \label{eqn:LinepackSoC}
\end{align}
\begin{align}
    lp_{k,l} = LP^{ini}_{l}
    \quad \forall~k\in\{1,K\},~l(m,n)\in\mathcal{L}
    \label{eqn:LinepackFinSoC}
\end{align}
Finally, inflow and outflow define the average pipeline gas flow $f_{k,l}$~\eqref{eqn:FlowAverage}.
\begin{align}
    f_{k,l} = \frac{f^{In}_{k,l} + f^{Out}_{k,l}}{2}
    \quad \forall~k\in\mathcal{K},~l(m,n)\in\mathcal{L}
    \label{eqn:FlowAverage}
\end{align}
The nodal pressures in~\eqref{eqn:Linepack} and the average gas flow in~\eqref{eqn:FlowAverage} are then related to the non-linear and non-convex general gas flow equation~\eqref{eqn:GasFlow} using a (piecewise) linearization method, e.g., INC, SOS2~\cite{Correa-Posada2014} or Z, which is proposed in Section~\ref{sec:Linearization}.

Moving gas along a pipeline requires a pressure difference.
To maintain adequate flow rates, compressor units are used in gas transmission systems to pressurize the gas. These units are typically installed in 100-200~km intervals and consume a percentage of the transported gas~\cite{Klatzer2022}. In the optimization model, compressors are represented in a simplified way, where the pressure of the incoming gas can be boosted by a constant compression ratio $\Lambda_{c}$~\eqref{eqn:Comp-Ratio-Rel} and its flow rate is limited by an upper bound $\overline{F}^{C}_{c}$, typically the maximum pipeline capacity~\eqref{eqn:Gas-BoundsCmp}.
\begin{align}
% Compressor ratio (relative) 
  p_{k,n} \leq \Lambda_{c} p_{k,m} 
  \quad \forall~k\in\mathcal{K},~c(m,n)\in\mathcal{C}
  \label{eqn:Comp-Ratio-Rel}
\end{align}
\begin{align}
% Bounds on compressor gas flow
  0 \leq f^{C}_{k,c} \leq \overline{F}^{C}_{c}
  \quad \forall~k\in\mathcal{K},~c(m,n)\in\mathcal{C}
  \label{eqn:Gas-BoundsCmp}
\end{align}
Finally, the nodal gas balance~\eqref{eqn:Gas-Balance} links gas production, demand and consumption of gas-fired power plants via pipeline gas flows. Note that a flow through a compressor increases the nodal gas demand by its consumption rate $CS^{G}_{c}$. 
\begin{align}
% Gas balance
  \sum_{s     \in\mathcal{A}^{\mathcal{S}}_m}      p^{G}_{k,s}             
 +\sum_{l(n,m)\in\mathcal{L}                }      f^{Out}_{k,l}        
 -\sum_{l(m,n)\in\mathcal{L}                }      f^{In}_{k,l}          
 +\sum_{c(n,m)\in\mathcal{C}                }      f^{C}_{k,c}    
 -\sum_{c(m,n)\in\mathcal{C}                }      f^{C}_{k,c}      
 +                                                 ns^{G}_{k,m}          \nonumber\\                   
 =
                                                   D^{G}_{k,m}               
 +\sum_{t\in\mathcal{A}^{\mathcal{T}}_m     }      cs^{G}_{k,t}             
 +\sum_{c(m,n)\in\mathcal{C}                }      CS^{G}_{c} f^{C}_{k,c} 
 \quad \forall~k\in\mathcal{K},~m\in\mathcal{M}
 \label{eqn:Gas-Balance}
\end{align}

\section{Proposed Linearization Method}
\label{sec:Linearization}
This section introduces the concept and mathematical framework of the proposed Z piecewise linearization method. In contrast to other piecewise methods, e.g., INC or SOS2~\cite{Correa-Posada2014}, Z linearizes the general gas flow equation~\eqref{eqn:GasFlow} based on the positive forward ($^+$)/reverse ($^-$) average gas flow $f^{+}_{k,l},~f^{-}_{k,l}$ and pressure difference $p^{+}_{k,l},~p^{-}_{k,l}$ respectively. 
Since Z is a self-contained, tight formulation, it can be utilized for quasi-dynamic gas flow modeling (this paper) or incorporated in dynamic gas flow modeling.

\subsection{Concept of Z Linearization}
\label{sec:ConceptZ}
\begin{figure*}
\centering
\includegraphics[scale=0.8]{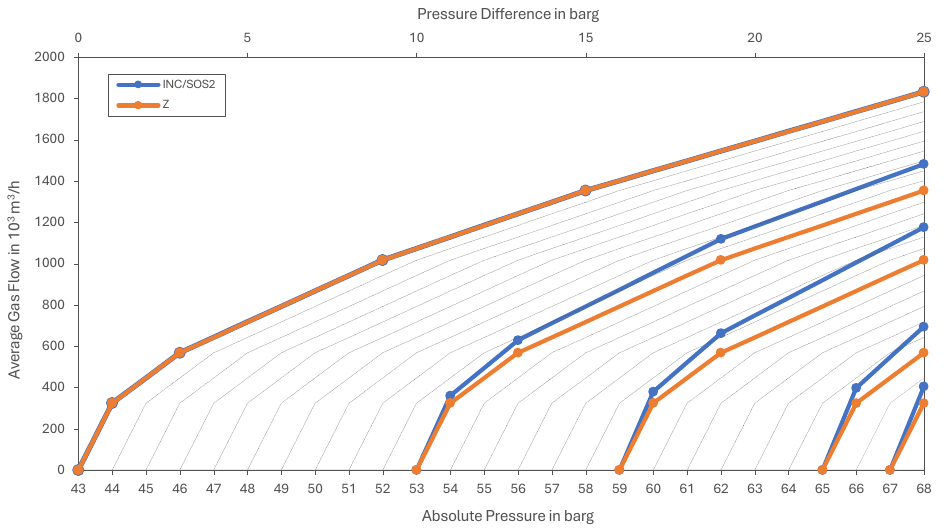}
\caption{Comparison of the linearized average gas flow under the Z and INC/SOS2 frameworks.}
\label{fig:Quality-Lin}
\end{figure*}

For the Z linearization,~\eqref{eqn:GasFlow} is rewritten as~\eqref{eqn:AvgGasFlowRe}. 
Here, the nodal pressure difference $(p_{k,m} - p_{k,n})$, which indicates the gas flow direction, is substituted by $(p^{+}_{k,l} + p^{-}_{k,l})$, where only one can take a value greater zero at a time.
\begin{align}
  f_{k,l} |f_{k,l}| = R^{G}_{l} (\underbrace{p_{k,m} - p_{k,n}}_\textit{\((p^{+}_{k,l} + p^{-}_{k,l})\)}) (p_{k,m} + p_{k,n}) 
  \quad \forall~k\in\mathcal{K},~l(m,n)\in\mathcal{L}
  \label{eqn:AvgGasFlowRe}
\end{align}
With this, $f_{k,l} |f_{k,l}|$ in~\eqref{eqn:AvgGasFlowRe} is expressed as squared average gas flow in~\eqref{eqn:GasPressRe}.
\begin{align}
  p^{+}_{k,l} + p^{-}_{k,l} = \frac{1}{R^{G}_{l}} \frac{f_{k,l}^2}{p_{k,m} + p_{k,n}} 
  \quad \forall~k\in\mathcal{K},~l(m,n)\in\mathcal{L}
  \label{eqn:GasPressRe}
\end{align}
We approximate~\eqref{eqn:GasPressRe} with the linear constraints~\eqref{eqn:PressDiffPWL}-\eqref{eqn:GasFlowPWL}, which relate and piecewise linearize $p^{+}_{k,l}, p^{-}_{k,l}$ and $f^{+}_{k,l}, f^{-}_{k,l}$ based on the parameters~$F_{l,z}$ and $P_{l,z}$ using the continuous variable~$\gamma_{k,l,z}$~\eqref{eqn:LOUP_gamma}. Parameter~$P_{l,z}$ is composed of the minimum pipeline pressure~$\underline{P}_m$, the pressure difference~$P^{\Delta}_{l,z}$, the corresponding value for the squared flow~$F^2_{l,z}$, and the pipeline parameter~$R^G_l$.
\begin{align}
  p^{+}_{k,l} + p^{-}_{k,l} = \sum_{z\in\mathcal{Z}} \underbrace{\frac{1}{R^{G}_{l}} \frac{F_{l,z}^2}{2\underline{P}_m+P^{\Delta}_{l,z}}}_\textit{$P_{l,z}$} \gamma_{k,l,z} 
  \quad \forall~k\in\mathcal{K},~l(m,n)\in\mathcal{L} 
  \label{eqn:PressDiffPWL}
\end{align}
\begin{align}
  f^{+}_{k,l} + f^{-}_{k,l} = \sum_{z\in\mathcal{Z}} F_{l,z} \gamma_{k,l,z} 
  \quad \forall~k\in\mathcal{K},~l(m,n)\in\mathcal{L}
  \label{eqn:GasFlowPWL}
\end{align}
As for the key logic of the Z piecewise linearization, the binary variable $\delta_{k,l,z}$~\eqref{eqn:Def_delta} ensures only adjacent~$\gamma_{k,l,z}$ take non-zero values~\eqref{eqn:SOS2Logic}, while \eqref{eqn:Sum_gamma}-\eqref{eqn:Sum_delta} force both their sums to equal~1. The binary variable~$\xi_{k,l}$~\eqref{eqn:Def_xi} determines the direction of the pressure difference~\eqref{eqn:PressPosDir}-\eqref{eqn:PressNegDir} and gas flow~\eqref{eqn:FlowPosDir}-\eqref{eqn:FlowNegDir} along a pipeline. Pressure difference and nodal pressures are related~\eqref{eqn:PressRelation} and bound~\eqref{eqn:LOUP_PressDiff}-\eqref{eqn:LOUP_NodalPress}. Similarly, gas flow and directional gas flow are related~\eqref{eqn:FlowRelation} and bound~\eqref{eqn:LOUP_Flow}-\eqref{eqn:LOUP_DirFlow}.
\begin{align}
  0 \leq \gamma_{k,l,z} \leq 1
  \quad \forall~k\in\mathcal{K},~l(m,n)\in\mathcal{L},~z\in\mathcal{Z}
  \label{eqn:LOUP_gamma}
\end{align}
\begin{align}
  \gamma_{k,l,z} \leq \delta_{k,l,z} + \delta_{k,l,z-1}
  \quad \forall~k\in\mathcal{K},~l(m,n)\in\mathcal{L},~z\in\mathcal{Z}
  \label{eqn:SOS2Logic}
\end{align}
\begin{align}
  \sum_{z\in\mathcal{Z}} \gamma_{k,l,z} = 1
  \quad \forall~k\in\mathcal{K},~l(m,n)\in\mathcal{L}
  \label{eqn:Sum_gamma}
\end{align}
\begin{align}
  \sum_{z\in\mathcal{Z}} \delta_{k,l,z} = 1
  \quad \forall~k\in\mathcal{K},~l(m,n)\in\mathcal{L}
  \label{eqn:Sum_delta}
\end{align}
\begin{align}
  0 \leq p^{+}_{k,l} \leq \xi_{k,l} \overline{P}_{l}
  \quad \forall~k\in\mathcal{K},~l(m,n)\in\mathcal{L}
  \label{eqn:PressPosDir}
\end{align}
\begin{align}
  0 \leq p^{-}_{k,l} \leq (1 - \xi_{k,l}) \overline{P}_{l}
  \quad \forall~k\in\mathcal{K},~l(m,n)\in\mathcal{L}
  \label{eqn:PressNegDir}
\end{align}
\begin{align}
  0 \leq f^{+}_{k,l} \leq \xi_{k,l} \overline{F}_{l}
  \quad \forall~k\in\mathcal{K},~l(m,n)\in\mathcal{L}
  \label{eqn:FlowPosDir}
\end{align}
\begin{align}
  0 \leq f^{-}_{k,l} \leq (1 - \xi_{k,l}) \overline{F}_{l}
  \quad \forall~k\in\mathcal{K},~l(m,n)\in\mathcal{L}
  \label{eqn:FlowNegDir}
\end{align}
\begin{align}
  p_{k,m} - p_{k,n} = p^{+}_{k,l} - p^{-}_{k,l}
  \quad \forall~k\in\mathcal{K},~l(m,n)\in\mathcal{L}
  \label{eqn:PressRelation}
\end{align}
\begin{align}
  0 \leq p^{+}_{k,l}, p^{-}_{k,l} \leq \overline{P}_{l}
  \quad \forall~k\in\mathcal{K},~l(m,n)\in\mathcal{L}
  \label{eqn:LOUP_PressDiff}
\end{align}
\begin{align}
  \underline{P}_{m} \leq p_{k,m} \leq \overline{P}_{m}
  \quad \forall~k\in\mathcal{K},~m\in\mathcal{M}
  \label{eqn:LOUP_NodalPress}
\end{align}
\begin{align}
  f_{k,l} = f^{+}_{k,l} - f^{-}_{k,l}
  \quad \forall~k\in\mathcal{K},~l(m,n)\in\mathcal{L}
  \label{eqn:FlowRelation}
\end{align}
\begin{align}
  -\overline{F}_{l} \leq f_{k,l} \leq \overline{F}_{l}
  \quad \forall~k\in\mathcal{K},~l(m,n)\in\mathcal{L}
  \label{eqn:LOUP_Flow}
\end{align}
\begin{align}
  0 \leq f^{+}_{k,l}, f^{-}_{k,l} \leq \overline{F}_{l}
  \quad \forall~k\in\mathcal{K},~l(m,n)\in\mathcal{L}
  \label{eqn:LOUP_DirFlow}
\end{align}
\begin{align}
  \delta_{k,l,z}\in\{0,1\} \quad \forall~k\in\mathcal{K},~l(m,n)\in\mathcal{L},~z\in\mathcal{Z}
  \label{eqn:Def_delta}
\end{align}
\begin{align}
  \xi_{k,l}\in\{0,1\} \quad \forall~k\in\mathcal{K},~l(m,n)\in\mathcal{L}
  \label{eqn:Def_xi}
\end{align}
\begin{align}
  0 \leq \delta_{k,l,z} \leq 1
  \quad \forall~k\in\mathcal{K},~l(m,n)\in\mathcal{L},~z\in\mathcal{Z}
  \label{eqn:LOUP_Delta_cont}
\end{align}
\begin{align}
  0 \leq \xi_{k,l} \leq 1
  \quad \forall~k\in\mathcal{K},~l(m,n)\in\mathcal{L}
  \label{eqn:LOUP_xi}
\end{align}
Fig.~\ref{fig:Quality-Lin} illustrates the approximation accuracy of the INC, SOS2 and Z piecewise linearization methods for a 1000~mm diameter pipeline over the full pressure range from 43 to 68~barg\footnote{This is based on the Austrian transmission system.}. The linearizations are based on selected pressure differences of 1, 3, 9, 15 and 25~barg and the evaluation of~\eqref{eqn:GasPressRe} for INC and SOS2 (in terms of absolute pressure), and~\eqref{eqn:PressDiffPWL} for Z. All linearizations assume an average gas velocity of 7~m/s. The main factors affecting the approximation accuracy are the number of linear segments used and the pressure range to be linearized. In this example, the full pressure range is linearized. Thus, the maximum relative approximation error is \textminus19.7\% (at $p_{k,m}=68$~barg; $p_{k,n}=67$~barg), making the Z linearization a more conservative approximation than INC/SOS2 at high absolute pressures. Generally, the approximation error can be limited by selecting appropriate pressure linearization points, e.g., derived from knowledge of typical operating conditions.
For example, choosing three linearization points between $p_{k,n}=53$~barg and $p_{k,m}=62$~barg, results in maximum approximation errors in the average gas flow of +4.7\% and \textminus4.3\%, respectively, for an absolute deviation in the pressure range of $62\pm5$~barg compared to INC/SOS2. However, the LP feasible region of constraints~\eqref{eqn:PressDiffPWL}\nobreakdash-\eqref{eqn:Def_xi} is not tight. The following section illustrates the process for deriving its tight feasible region~$\mathcal{H}^\mathrm{t}$.

\subsection{Deriving Tight Feasible Region}
\label{sec:PORTA}
The LP feasible region~$\mathcal{N}$ of the full set of mixed-integer linear constraints \eqref{eqn:PressDiffPWL}\nobreakdash-\eqref{eqn:Def_xi} is not tight, e.g., because it contains big-M constraints, which are not tight by definition (see Section~\ref{sec:Intro}). Figure~\ref{fig:FlowChart} illustrates the process for deriving the tight feasible region~$\mathcal{H}^\mathrm{t}$ from~$\mathcal{N}$ using the Polyhedron Representation Transformation Algorithm (PORTA) software package (v1.4.1)~\cite{PORTA}.
In PORTA, for a given set of input parameters~$j\in\mathcal{J}$, e.g., $\overline{F}_{l,j}, P_{l,z,j}$, \eqref{eqn:PressDiffPWL}\nobreakdash-\eqref{eqn:Def_xi} are represented by their equivalent relaxed set of constraints~\eqref{eqn:PressDiffPWL}-\eqref{eqn:LOUP_DirFlow} and~\eqref{eqn:LOUP_Delta_cont}-\eqref{eqn:LOUP_xi} with feasible region~$\mathcal{H}_j$. PORTA transforms~$\mathcal{H}_j$ from the constraint space~(CS) to the vertex space~(VS). In the VS, the feasible region~$\mathcal{V}_j$ is tightened by adding strong valid inequalities, i.e., cuts, that satisfy integrality of binary variables in the relaxed problem, which results in a set of tight vertices~$\mathcal{V}^\mathrm{t}_j$ in the VS. Then, the set of tight vertices is transformed back to the CS, corresponding to a linear system of constraints with the tight feasible region~$\mathcal{H}^\mathrm{t}_j$. 
Finally, the linear constraints from PORTA representing~$\mathcal{H}^\mathrm{t}_j$ are interpreted and generalized to the set of linear constraints that model the tight feasible region~$\mathcal{H}^\mathrm{t}$, which follows in the next section.
\begin{NoHyper}
\begin{figure}[!h]
\centering
\begin{tikzpicture}[scale=0.6, every node/.style={transform shape}]
% costom hex colors
\definecolor{green}{HTML}{A1E3A1}
\definecolor{orange}{HTML}{F8CBAD}
\definecolor{blue}{HTML}{A3C1DA}
\definecolor{yellow}{HTML}{FFECB3}

\tikzstyle{MI} = [rectangle, 
minimum width=4cm, 
minimum height=2.5cm,
text centered,
text width=4cm,
draw=gray, 
fill=gray!10]

\tikzstyle{Rel} = [rectangle, 
minimum width=4cm, 
minimum height=2.5cm,
text centered,
text width=4cm,
draw=gray, 
fill=gray!10]

\tikzstyle{LSC} = [rectangle, 
minimum width=4cm, 
minimum height=2.5cm,
text centered,
text width=4cm,
draw=orange, 
fill=orange!30]

\tikzstyle{VS} = [rectangle, 
minimum width=4cm, 
minimum height=2.5cm,
text centered,
text width=4cm,
draw=blue, 
fill=blue!30]

\tikzstyle{arrow} = [thick,->,>=stealth]

% Positioning of nodes
\node (0)   [MI,              xshift=+0.0cm, yshift=+0.0cm] {\textbf{Mixed-integer \\ linear constraints} \\ \eqref{eqn:PressDiffPWL}--\eqref{eqn:Def_xi} with LP \\ feasible region $\mathcal{N}$};
 
\node (1)   [Rel, right of=0, xshift=+4.2cm, yshift=+0.2cm] {\textbf{Relaxed \\ constraints} \\ \eqref{eqn:PressDiffPWL}--\eqref{eqn:LOUP_DirFlow}, \eqref{eqn:LOUP_Delta_cont}--\eqref{eqn:LOUP_xi} \\ for input parameters $j\in\mathcal{J}$};
\node (2)   [Rel, right of=0, xshift=+4.1cm, yshift=+0.1cm] {\textbf{Relaxed \\ constraints} \\ \eqref{eqn:PressDiffPWL}--\eqref{eqn:LOUP_DirFlow}, \eqref{eqn:LOUP_Delta_cont}--\eqref{eqn:LOUP_xi} \\ for input parameters $j\in\mathcal{J}$};
\node (3)   [Rel, right of=0, xshift=+4.0cm, yshift=+0.0cm] {\textbf{Relaxed \\ constraints} \\ \eqref{eqn:PressDiffPWL}--\eqref{eqn:LOUP_DirFlow}, \eqref{eqn:LOUP_Delta_cont}--\eqref{eqn:LOUP_xi} \\ for input parameters $j\in\mathcal{J}$};
 
\node (4)   [LSC, right of=1, xshift=+4.0cm, yshift=+0.0cm] {\textbf{PORTA} (CS) \\ Feasible region $\mathcal{H}_3$};
\node (5)   [LSC, right of=2, xshift=+4.0cm, yshift=+0.0cm] {\textbf{PORTA} (CS) \\ Feasible region $\mathcal{H}_2$};
\node (6)   [LSC, right of=3, xshift=+4.0cm, yshift=+0.0cm] {\textbf{PORTA} (CS) \\ Feasible region $\mathcal{H}_j$};
 
\node (7)   [VS, right of=4,  xshift=+5.8cm, yshift=+0.0cm] {\textbf{PORTA} (VS) \\ Set of vertices $\mathcal{V}_3$};
\node (8)   [VS, right of=5,  xshift=+5.8cm, yshift=+0.0cm] {\textbf{PORTA} (VS) \\ Set of vertices $\mathcal{V}_2$};
\node (9)   [VS, right of=6,  xshift=+5.8cm, yshift=+0.0cm] {\textbf{PORTA} (VS) \\ Set of vertices $\mathcal{V}_j$};
 
\node (10)  [VS, below of=7,  xshift=+0.0cm, yshift=-2.5cm] {\textbf{PORTA} (VS) \\ Set of tight vertices $\mathcal{V}^\mathrm{t}_3$};
\node (11)  [VS, below of=8,  xshift=+0.0cm, yshift=-2.5cm] {\textbf{PORTA} (VS) \\ Set of tight vertices $\mathcal{V}^\mathrm{t}_2$};
\node (12)  [VS, below of=9,  xshift=+0.0cm, yshift=-2.5cm] {\textbf{PORTA} (VS) \\ Set of tight vertices $\mathcal{V}^\mathrm{t}_j$};

\node (13)  [LSC, below of=4, xshift=+0.0cm, yshift=-2.5cm] {\textbf{PORTA} (CS) \\ Tight feasible region $\mathcal{H}_3^\mathrm{t}$};
\node (14)  [LSC, below of=5, xshift=+0.0cm, yshift=-2.5cm] {\textbf{PORTA} (CS) \\ Tight feasible region $\mathcal{H}_2^\mathrm{t}$};
\node (15)  [LSC, below of=6, xshift=+0.0cm, yshift=-2.5cm] {\textbf{PORTA} (CS) \\ Tight feasible region $\mathcal{H}_j^\mathrm{t}$};

\node (16)  [MI,  below of=0, xshift=+0.0cm, yshift=-2.5cm] {\textbf{Linear constraints} \\ \eqref{eqn:EQ3}--\eqref{eqn:IEQ6} with tight feasible region $\mathcal{H}^\mathrm{t}$};

% Definition of arrows connecting nodes
\draw [arrow] (0)  -- node[anchor=south ] {} (3);
\draw [arrow] (3)  -- node[anchor=south ] {} (6);
\draw [arrow] (6)  -- node[anchor=south ] {Transform} (9);
\draw [arrow] (9)  -- node[anchor=west  ] {Tighten} (12); 
\draw [arrow] (12) -- node[anchor=south ] {Transform} (15);
\draw [arrow] (15) -- node[anchor=center, align=center] {Interpret \\ \& Generalize} (16);

\draw [arrow, gray] ([xshift=-0.2cm, yshift=-0.05cm]3.north west) -- ([xshift=0.00cm, yshift=0.2cm]1.north west)  node[midway, text=black, sloped, yshift=+0.25cm, ] {\footnotesize $j\in\mathcal{J}$};
\end{tikzpicture}
\caption{Process for deriving tight feasible region with PORTA.}
\label{fig:FlowChart}
\end{figure}
\end{NoHyper}

\subsection{Set of Tight Constraints}
\label{sec:TightConstraints}
This section presents the generalized set of linear constraint~\eqref{eqn:EQ3}\nobreakdash-\eqref{eqn:IEQ6} that describe the tight feasible region~$\mathcal{H}^\mathrm{t}$.
Fundamentally,~\eqref{eqn:EQ3}\nobreakdash-\eqref{eqn:IEQ6} follow the same concept as~\eqref{eqn:PressDiffPWL}\nobreakdash-\eqref{eqn:Def_xi}. However, as a direct result of the tightening in PORTA,~\eqref{eqn:EQ3}\nobreakdash-\eqref{eqn:IEQ6} are designed in such a way that binary variables~$\delta_{k,l,z}$ and~$\xi_{k,l}$ always take binary values, even when integrality is relaxed. This is due to the cuts added in the vertex space in PORTA, which ensures that relaxed binary variables only take binary values in the vertices.

Achieving a tight feasible region requires calculating parameters $\Tilde{A}_{l,z,\Tilde{z}}$ to $\Tilde{F}_{l,u,v,z}$, $Z^{sgn}_{l,u,v,w}$ and $Z^{RHS}_{l,u,v,w}$. The respective pre-processing algorithm for calculating these parameters is presented in~\ref{sec:Appendix}. 
Note that for these calculations, $P_{l,z}$ and $F_{l,z}$ must be integer. This is to determine discrete combinations of the indices $u,v,w$ via the pre-processing algorithm in order to formulate the set of tight constraints. For the case study in this paper, the corresponding computational time of the algorithm is approximately 80 to 200 seconds per pipeline. However, these calculations can be parallelized. \\

Identical to the set of constraints presented in Section~\ref{sec:ConceptZ}, in the set of tight constraints the gas flow is expressed as directional gas flows~\eqref{eqn:EQ3} and the nodal pressures are expressed as pressure differences~\eqref{eqn:EQ4}. The directional gas flow and pressure difference are related and piecewise linearized in~\eqref{eqn:EQ2}\nobreakdash-\eqref{eqn:EQ1}.
\begin{align}
    f_{k,l} = f^{+}_{k,l} - f^{-}_{k,l}
    \quad \forall~k\in\mathcal{K},~l(m,n)\in\mathcal{L}
    \label{eqn:EQ3}
\end{align}
\begin{align}
    p_{k,m} - p_{k,n} = p^{+}_{k,l} - p^{-}_{k,l}
    \quad \forall~k\in\mathcal{K},~l(m,n)\in\mathcal{L}
\label{eqn:EQ4}
\end{align}
\begin{align}
    p^{+}_{k,l} + p^{-}_{k,l} = \sum_{z\in\mathcal{Z}} P_{l,z} \gamma_{k,l,z}
    \quad \forall~k\in\mathcal{K},~l(m,n)\in\mathcal{L}
\label{eqn:EQ2}
\end{align}
\begin{align}
    P_{l,z=Z} \sum_{z\in\{1,\dots,Z-1\}} F_{l,z} \gamma_{k,l,z}
  - F_{l,z=Z} \sum_{z\in\{1,\dots,Z-1\}} P_{l,z} \gamma_{k,l,z} \nonumber \\
  =
    P_{l,z=Z} (f^{+}_{k,l} + f^{-}_{k,l})
  - F_{l,z=Z} (p^{+}_{k,l} + p^{-}_{k,l}) 
    \quad \forall~k\in\mathcal{K},~l(m,n)\in\mathcal{L}  
\label{eqn:EQ1}
\end{align}
The key logic of the original piecewise linearization -- that only adjacent~$\gamma_{k,l,z}$ take non-zero values -- remains unchanged. However, to achieve a tight feasible region, original constraints~\eqref{eqn:LOUP_gamma}-\eqref{eqn:Sum_delta} and~\eqref{eqn:Def_delta}-\eqref{eqn:Def_xi} are expressed as~\eqref{eqn:EQ5}-\eqref{eqn:IEQ2}. Again, tightness of the feasible region forces~$\delta_{k,l,z}$ to always take binary values, even when defined as continuous variable.
\begin{align}
    \sum_{z\in\mathcal{Z}} \gamma_{k,l,z} = 1
    \quad \forall~k\in\mathcal{K},~l(m,n)\in\mathcal{L} 
\label{eqn:EQ5}
\end{align}
\begin{align}
    \sum_{z\in\mathcal{Z}} \delta_{k,l,z} = 1
    \quad \forall~k\in\mathcal{K},~l(m,n)\in\mathcal{L}
\label{eqn:EQ6}
\end{align}
\begin{align}
    \sum_{z\in\{1,\dots,Z-1\}} \gamma_{k,l,z}
    \leq
    1            
    \quad \forall~k\in\mathcal{K},~l(m,n)\in\mathcal{L},\ {\text{if}}\ Z>3
\label{eqn:IEQ7}
\end{align}
\begin{align}
    \sum_{z < \Tilde{z} \leq Z} \delta_{k,l,\Tilde{z}}
    \leq
    \sum_{z < \Tilde{z} \leq Z} \gamma_{k,l,\Tilde{z}}
    \quad \forall~k\in\mathcal{K},~l(m,n)\in\mathcal{L},~z\in\{1,\dots,Z-1\}
\label{eqn:IEQ1}
\end{align}
\begin{align}
    \sum_{z <   \Tilde{z} \leq Z} \delta_{k,l,\Tilde{z}}
    \geq
    \sum_{z+1 < \Tilde{z} \leq Z} \gamma_{k,l,\Tilde{z}}
    \quad \forall~k\in\mathcal{K},~l(m,n)\in\mathcal{L},~z\in\{1,\dots,Z-1\}
\label{eqn:IEQ2}
\end{align}
Constraints~\eqref{eqn:IEQ4}-\eqref{eqn:IEQ3} establish tight lower and upper bounds on feasible combinations of the reverse average gas flow and the corresponding pressure difference simultaneously. Note that their structure is a design choice to achieve tight inequality constraints.
\begin{align}
    P_{l,z} f^{-}_{k,l} 
  - F_{l,z} p^{-}_{k,l}
    \geq
    \sum_{\Tilde{z}\in\mathcal{Z}} \Tilde{A}_{l,z,\Tilde{z}} \gamma_{k,l,\Tilde{z}}   
    \quad \forall~k\in\mathcal{K},~l(m,n)\in\mathcal{L},~z\in\{2,\dots,Z\}          
\label{eqn:IEQ4}
\end{align}
\begin{align}
    P_{l,z} f^{-}_{k,l} 
  - F_{l,z} p^{-}_{k,l}
    \leq
    \sum_{\Tilde{z}\in\mathcal{Z}} \Tilde{B}_{l,\Tilde{z},z} \gamma_{k,l,\Tilde{z}}
  - \sum_{\Tilde{z}\in\mathcal{Z}} \Tilde{C}_{l,z,\Tilde{z}} \gamma_{k,l,\Tilde{z}}    \nonumber \\
    \quad \forall~k\in\mathcal{K},~l(m,n)\in\mathcal{L},~z\in\{2,\dots,Z\}  
\label{eqn:IEQ3}
\end{align}
Constraints~\eqref{eqn:IEQ8}-\eqref{eqn:IEQ9} determine the gas flow direction via~$\xi_{k,l}$.
Here, indices~$u\in\mathcal{U}=\{1,2,\dots,max(F_{l,z=Z}\!-\!F_{l,z=2}, P_{l,z=Z}\!-\!P_{l,z=2})\}$ with alias $v$, and $w\in\mathcal{W} = \{1,2,\dots,\lfloor\frac{Z-1}{2} \rfloor \}$ are introduced. The pre-processing algorithm determines valid combinations of $u,v,w$ (for which $P_{l,z}$ and $F_{l,z}$ must be integer) and calculates the corresponding parameters $\Tilde{D}_{l,u,v,z}, \Tilde{E}_{l,u,v,z}, \Tilde{F}_{l,u,v,z}, Z^{RHS}_{l,u,v,w}$ and $Z^{sgn}_{l,u,v,w}$ to establish tight inequalities. Again, tightness of the constraints forces~$\xi_{k,l}$ to take binary values, even when defined as a continuous variable. Note that the parameters $u$ and $v$ correspond to the order of the determined valid combinations of $u$ and $v$ indices. In terms of notation, $sgn(\cdot)$ is the signum function.
\begin{align}
  - sgn(Z^{sgn}_{l,u,v,w}) u \ f^{-}_{k,l}
  + sgn(Z^{sgn}_{l,u,v,w}) v \ p^{-}_{k,l}                          \nonumber \\
  + \sum_{z\in\{2,\dots,Z\}} (\Tilde{D}_{l,u,v,z} + \Tilde{E}_{l,u,v,z}) \gamma_{k,l,z}
 \leq
  - \xi_{k,l} Z^{RHS}_{l,u,v,w}                                     \nonumber \\
    \quad \forall~k\in\mathcal{K},~l(m,n)\in\mathcal{L}~\cap~Z^{RHS}_{l,u,v,w} \neq 0        
\label{eqn:IEQ8}
\end{align}
\begin{align}
  + sgn(Z^{sgn}_{l,u,v,w}) u \ f^{-}_{k,l}
  - sgn(Z^{sgn}_{l,u,v,w}) v \ p^{-}_{k,l}                          \nonumber \\
  + \sum_{z\in\{2,\dots,Z\}} \Tilde{F}_{l,u,v,z} \gamma_{k,l,z}
 \leq
  - Z^{RHS}_{l,u,v,w} (1 - \xi_{k,l})                               \nonumber \\
    \quad \forall~k\in\mathcal{K},~l(m,n)\in\mathcal{L}~\cap~Z^{RHS}_{l,u,v,w} \neq 0        
\label{eqn:IEQ9}
\end{align}
Finally,~\eqref{eqn:IEQ5}-\eqref{eqn:IEQ6} establish tight lower and upper bounds on the nodal pressures\footnote{Note that the bounds on $p^+_{k,l}$ and $p_{k,m}$ are implicitly included in the set of tight constraints~\eqref{eqn:EQ3}--\eqref{eqn:IEQ6}.}.
\begin{align}
    \underline{P}_m \leq p_{k,n} - p^{-}_{k,l}                 
    \quad \forall~k\in\mathcal{K},~l(m,n)\in\mathcal{L}
\label{eqn:IEQ5}
\end{align}
\begin{align}
    p_{k,n} - p^{-}_{k,l}
    \leq
    \overline{P}_m - \sum_{z\in\{2,\dots,Z\}} P_{l,z} \gamma_{k,l,z}  
    \quad \forall~k\in\mathcal{K},~l(m,n)\in\mathcal{L}
\label{eqn:IEQ6}
\end{align}

\section{Numerical Results}
\label{sec:Numerical}
In this section, we illustrate the characteristics and study the performance of Z compared with the INC and SOS2 piecewise linearization methods. First, we compare their tightness and compactness, which gives a theoretical indication about their expected computational performance. Then we apply the formulations in various case studies of an integrated power and gas ESOM and analyze their operational results and practical computational performance.

\subsection{Case Study Setup}
\label{sec:CaseStudySetup}
\begin{figure*}[!t]
\centering
\includegraphics[scale=0.52]{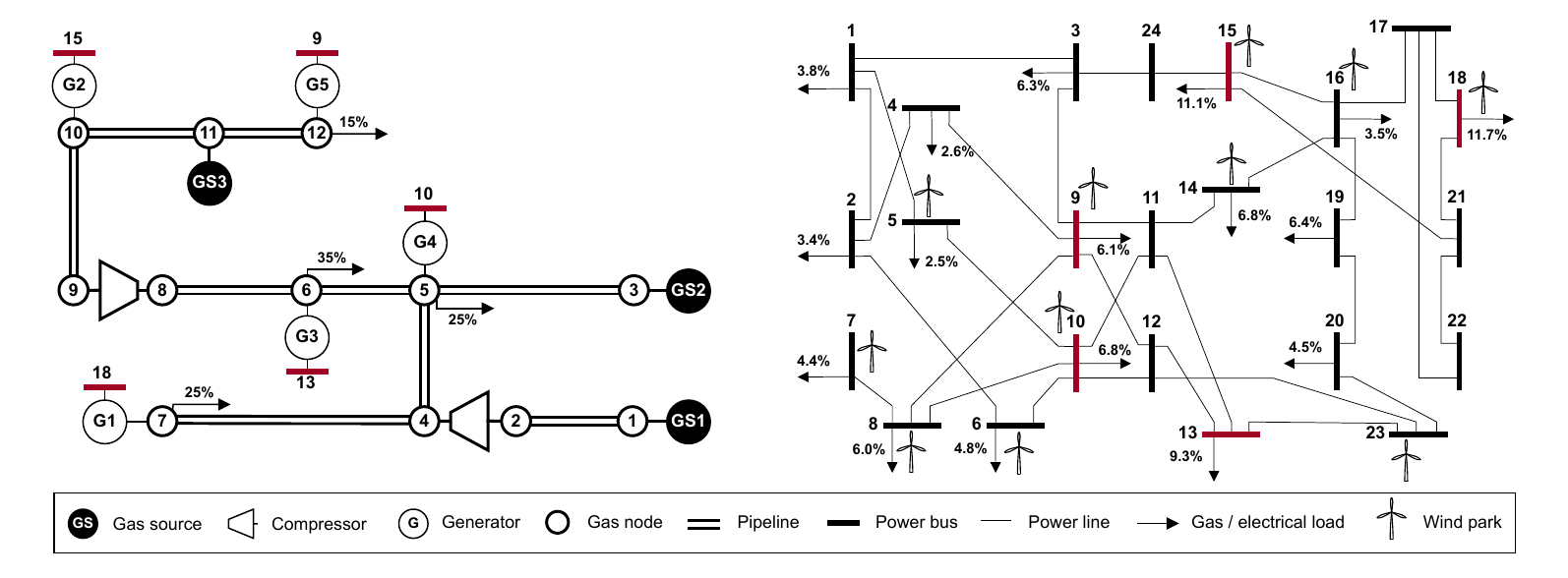}
\caption{Integrated power and gas system.}
\label{fig:ES}
\end{figure*}

The case study is based on a modified version of an integrated 24-bus IEEE Reliability Test System and a 12-node gas system depicted in Fig.~\ref{fig:ES}~\cite{Klatzer2022}. The systems are interlinked by 5 gas-fired generation units with a total capacity of 1,476~MW. The installed renewable capacity amounts to 6,000~MW wind and 3,130~MW solar and is modeled based on capacity factors. Table~\ref{tab:GenData} provides the techno-economic data for the generation units. 
For the high-pressure gas transmission system, we assume a uniform inner pipeline diameter of 800~mm and a typical operating pressure of 43--68~barg (based on the Austrian system). 
Table~\ref{tab:GasSysData} summarizes the most relevant data of the pipeline system. 
Power flows are approximated by a DC-optimal power flow based on voltage angles. 
The power and gas demands follow the downscaled Austrian time series (hourly resolution) and are distributed across gas nodes and power busses.
All case studies are solved on a Workstation with a 12\textsuperscript{th} Generation Intel Core i9-12900 (2.4~GHz, 16~cores, 24~threads) with 128~GB RAM using GAMS 46.4.1 and Gurobi 11.0.1. with default settings. The data is accessible in a GitHub repository~\cite{GitHubRepo_DataPre_2024}.

\subsection{Results}
\label{sec:Results}
\subsubsection{Tightness}
\label{sec:Tight}
The tightness of a MILP model is measured by the integrality gap~\cite{Morales2013}, defined as the difference between the objective function values of the MILP model and its LP relaxation. However, explicitly measuring the tightness of a subset of constraints in a MILP model is non-trivial, as it is generally impacted by linking constraints from the subset with constraints outside the subset, e.g., $f_{k,l}$ links~\eqref{eqn:FlowAverage} and~\eqref{eqn:EQ3}.
In order to analyze the tightness of the INC, SOS2 and Z formulations without interference of other constraints, which is how different MILP formulation are usually compared~\cite{Vielma2010}, we compute the number of vertices of their respective relaxed LP feasible regions using PORTA. In this context, we can evaluate their tightness by comparing the number of binary variables in the polyhedron that take fractional values, while the number of vertices in the polyhedron is a proxy for the LP complexity.
Table~\ref{tab:Polyhedra} presents the results for a single pipeline for two sets of linear segments (a) and (b). These numbers clearly illustrate that applying INC and SOS2 to linearize both average gas flow and nodal pressures, i.e., their union, harms the tightness of the resulting MILP as more than 90\% of vertices include fractional binaries. Note that the number of vertices for INC are more than 42 and 247 times larger than for Z. For SOS2 case (a), this is more than 1009 times, while PORTA cannot compute SOS2 case (b). In contrast to INC and SOS2, the Z linearization has the characteristics of a tight formulation, in fact the tightest possible, at least for the parameters used for these case studies. 
\begin{table}[!t]
\footnotesize
\centering
\begin{tabular}{|c c c c c c|}
\hline
Gen        & $\overline{P}$ & $\underline{P}$   & $RU/RD$   & $C^{OM}$  & $CS^{V}$  \\
           & [MW]           & [MW]              & [MW]      & [\$/MWh]  & [MWh\textsubscript{th}/MWh]      \\
\hline
G1         & 480            & 280               & 200       & 4         & 1.96        \\
G2         & 404            & 260               & 144       & 4         & 1.79        \\
G3         & 240            & 116               & 124       & 4         & 2.17        \\
G4         & 240            & 116               & 124       & 4         & 2.17        \\
G5         & 112            & 77                & 35        & 4         & 2.57        \\
Wind       & 50             & -                 & -         & 2         & -           \\
Solar      & 5              & -                 & -         & -         & -           \\
\hline
\end{tabular}
\caption{Generator Data.\label{tab:GenData}}
\end{table}
\begin{table}[!t]
\footnotesize
\centering
\begin{tabular}{|c c c c c|}
\hline
Pipeline   & Length  & $R^{G}$   & $\overline{F}$   & LP        \\
           & [km]    & [(kSm\textsuperscript{3}/h barg)\textsuperscript{2}]        & [kSm\textsuperscript{3}/h]         & [kSm\textsuperscript{3}/barg]  \\
\hline
1--2, 3--5, 4--7     & 70       & 379.82       & 1026.65       & 42.84     \\
5--4, 5--6           & 60       & 443.13       & 1108.91       & 36.72     \\
6--8                 & 50       & 332.34       & 960.34        & 48.96     \\
9--10, 10--11        & 40       & 664.69       & 1358.13       & 24.48     \\
11--12               & 100      & 265.88       & 858.95        & 61.20     \\
\hline
\end{tabular}
\caption{High-Pressure Pipeline Transmission System Data.\label{tab:GasSysData}}
\end{table}
\begin{table*}[!t]
\footnotesize
\centering
\begin{threeparttable}
\begin{tabular}{|c c c c c c c|} 
\hline  
    & \multicolumn{2}{c}{INC} & \multicolumn{2}{c}{SOS2}  & \multicolumn{2}{c|}{Z}    \\
    & (a) & (b)               & (a) & (b)                 & (a) & (b)                 \\
\hline
\# vertices in polyhedron                 & 1,202 & 9,888  & 28,254 & * &   28 &   40 \\  
\% vertices with fractional binaries      & 94.34 & 96.78  &  99.53 & * & 0.00 & 0.00 \\
$\text{average}\Big(\frac{\text{\# fractional binaries}}{\text{\# fractional vertices}}\Big)$ &  2.73 &  4.18  &   4.51 & * & 0.00 & 0.00 \\
\hline
\end{tabular}
\begin{tablenotes}
    \item \noindent
    (a) INC/SOS2: 2x3 nodal pressure segments; ~6 average gas flow segments. Z: 3 segments.
    \item \noindent
    (b) INC/SOS2: 2x5 nodal pressure segments; 10 average gas flow segments. Z: 5 segments.
    \item \noindent
    * Cannot be computed by PORTA.
\end{tablenotes}
\end{threeparttable}
\caption{Characteristics of Polyhedra.\label{tab:Polyhedra}}
\end{table*}

\subsubsection{Problem Size}
\label{sec:ProbSize}
As mentioned in the introduction, the computational performance of an MILP model is not solely impacted by its tightness, but also by its compactness or model size. Typically, tightening can be achieved by adding tighter valid inequalities, i.e., cuts. However, increasing the number of constraints implies repeatedly solving larger LP relaxations during the branch and bound phase, which can worsen computational performance compared to a less tight model~\cite{Morales2013}.
Table~\ref{tab:ProblemSize} presents the number of constraints, continuous and binary variables, where $\mathcal{L}$ and $\mathcal{M}$ are the number of pipelines and gas nodes, $\mathcal{F}$ and $\mathcal{P}$ are the number of flow and pressure grid points for the INC and SOS2, and $\mathcal{Z}$ is the number of grid points for the Z linearization. In addition, the number of nonzeros is illustrated.
Numbers are shown for: a single pipeline (1P), the pipeline system (PS) and the complete energy system (ES) in Fig.~\ref{fig:ES}\footnote{Numbers are based on case (b). For 1P and PS the time horizon is one hour and the number of constraints, variables and nonzeros increases linearly with the number of time periods modeled. Numbers for ES refer to a model run with 24 time periods.}. As shown, the number of constraints and nonzeros, which are proxies for the compactness of a formulation, increases under Z, while the number of variables is reduced by 1.77 and 2.43 times for continuous and binary variables compared to INC. Consequently, Z is a more compact formulation with respect to the number of variables. However, given the above, it is difficult to make a priori statements about the expected computational performance of Z based on problem size. Rather, it is the combination of tightness, number of vertices in the polyhedron and problem size that determines computational performance in practice, which we study in the following section.
\begin{landscape}
\begin{table*}[!t]
\footnotesize
\centering
\begin{threeparttable}
\begin{tabular}{|c|c|c|c|c|c|}
\hline
\multicolumn{2}{|c|}{Case} & \# constraints & \# continuous variables & \# binary variables & \# nonzeros \\
\hline\hline
\multirow{4}{*}{INC} & -- & $2(\mathcal{L}\!+\!\mathcal{L}(\mathcal{F}\!-\!2)\!+\!\mathcal{M}(\mathcal{P}\!-\!2))\!+\!\mathcal{M}$ & $\mathcal{L}\!+\!\mathcal{M}\!+\!\mathcal{L}(\mathcal{F}\!-\!1)\!+\!\mathcal{M}(\mathcal{P}\!-\!1)$ & $\mathcal{L}(\mathcal{F}\!-\!2)\!+\!\mathcal{M}(\mathcal{P}\!-\!2)$ & -- \\ \cline{2-6}
    & 1P & 38     & 23     & 17    & 111    \\
    & PS & 288    & 171    & 129   & 835    \\ 
    & ES & 11,304 & 10,354 & 3,463 & 37,333 \\ \hline\hline
\multirow{4}{*}{SOS2} & -- & $3\mathcal{L}\!+\!2\mathcal{M}$ & $\mathcal{L}\!+\!\mathcal{M}\!+\!\mathcal{L}\mathcal{F}\!+\!\mathcal{M}\mathcal{P}$ & --*& -- \\ \cline{2-6}
    & 1P & 7      & 26     & *     & 70     \\ 
    & PS & 51     & 192    & *     & 552    \\ 
    & ES & 5,616  & 10,858 & 367*  & 30,541 \\ \hline\hline
\multirow{4}{*}{Z} & -- & $9\mathcal{L}\!+\!8\mathcal{L}(\mathcal{Z}\!-\!1)$ & $5\mathcal{L}\!+\!\mathcal{M}\!+\!\mathcal{L}\mathcal{Z}$ & $\mathcal{L}\!+\mathcal{L}\mathcal{Z}$ & -- \\ \cline{2-6}
    & 1P & 49     & 13    & 7     & 269    \\ 
    & PS & 421    & 111   & 63    & 2,421  \\ 
    & ES & 14,976 & 8,698 & 1,879 & 75,181 \\  
\hline
\end{tabular}
\begin{tablenotes}
    \item \noindent
    * The SOS2 formulation does not use explicit binaries, but SOS2-type variables, which are handled by the solver. Binaries reported for the SOS2-ES case are related to UC decisions.
\end{tablenotes}
\end{threeparttable}
\caption{Problem Size.\label{tab:ProblemSize}}
\end{table*}
\end{landscape}

\subsubsection{Computational Performance}
\label{sec:CompPerf}
In the following, we evaluate the computational performance of the ESOM under INC, SOS2 and Z for a representative summer (S-24) and winter (W\nobreakdash-24) day (hourly resolution) and conduct sensitivity analyses with respect to the number of piecewise linear segments, increased temporal complexity (S-48), gas losses and gas costs. Model statistics include the objective function value, the number of explored nodes, the relative MILP gap, CPU time and the resulting speed-up of Z over INC or SOS2, depending on which has the shorter CPU time. 
To increase the level of computational complexity, gas-fired generators are modeled based on discrete unit commitment decisions~\eqref{eqn:UCTotOut}-\eqref{eqn:UCConversionP} and the problem is solved to a relative MILP optimality gap of 0.1\%.

Table~\ref{tab:SvsW-diff} shows the model statistics for three sets of piecewise linear segments (a), (b) and (c) assuming a moderate difference in gas costs per source (GS1:~+0.8\%; GS2:~+0.0\%; GS3:~+0.4\%). All cases yield the same objective function value for Z and INC at similar final gaps -- independent of the number of segments used. Note that under SOS2, none of the case can be solved to the optimality gap within the 3600-second time limit. 
The number of explored nodes gives an indication of the complexity of the LP relaxations during the solution procedure~\cite{Morales2013}. Here, Z explores on average approximately twice the number of nodes per unit of time compared to INC. Consequently, there is a significant difference in the CPU time for the model solution, with an average speed-up of 3.72 for the summer (lower gas and power demand) and 2.18 for the winter days (high gas and power demand). Note that despite using more segments, case S-24 (b) shows faster CPU times than case (a), which seems counterintuitive but is consistent under both INC and Z. Potential reasons for this could be the different feasible region when using more segments or the use of heuristics during branch and bound under Gurobi's default settings.
For the winter case, both the number of explored nodes and the CPU time increase.
This is mainly due to the combination of high gas demand and temporal linking established by linepack, where the ESOM tries to minimize operational cost by storing gas from the least-cost gas well for later use. Note that INC performs better for case W-24 (c) than Z, which does not solve to the gap of 0.1\% within the time limit. For increased temporal complexity (48 hours), only case S-48 (b) can be solved.

Table~\ref{tab:Loss} illustrates the impact of gas losses on model statistics for case W-24 (a), again for a moderate difference in gas cost. In general, considering gas losses in the linepack state of charge formulation~\eqref{eqn:LinepackSoC} results in longer CPU times, with INC eventually exceeding the solving time limit of one hour. Moreover, an opposing trend between INC and Z can be observed for increasing losses.

Finally, Table~\ref{tab:GasCost} shows the model statistics for identical and very different gas costs (GS1:~+84\%; GS2:~+0.0\%; GS3:~+42\%) under cases S-24 (a) and W-24 (a), which are ambiguous. In the latter case, the summer day stands out, as INC performs better than Z (and even the base case in Table~\ref{tab:SvsW-diff}). However, as observed before, for the winter day, both the number of nodes explored and CPU time increase, which is amplified by the difference in gas cost. We conclude that one cannot make statements a priori on computational performance based on a high difference in gas cost, although this may initially seem intuitive. 
In the case of identical gas costs, it appears that INC benefits from "symmetry" in the optimization problem, as both summer and winter days show similar or improved CPU times compared to the base case, while the opposite is true for Z.
\begin{landscape}
\begin{table*}[!t]
\footnotesize
\centering
\begin{threeparttable}
\begin{tabular}{|c c|c c c|c c c|c c c|c c c|c|}
\hline
    \multicolumn{2}{|c|}{Case}          & \multicolumn{3}{c|}{Objective Function}    & \multicolumn{3}{c|}{\# Nodes $[\times 10^3]$} & \multicolumn{3}{c|}{Gap}     & \multicolumn{3}{c|}{CPU Time}     & Speed-up  \\ 
    \multicolumn{2}{|c|}{}              & \multicolumn{3}{c|}{[M\$]}                 & \multicolumn{3}{c|}{[--]}                     & \multicolumn{3}{c|}{[\%]}    & \multicolumn{3}{c|}{[s]}          & [--]      \\ \cline{3-15}
    ~                          &  ~     & INC   & SOS2   & Z                         & INC    & SOS2  & Z                            & INC   & SOS2  & Z            & INC   & SOS2 & Z                  & --        \\ \hline
    \multirow{3}{*}{S-24}      & (a)    & 2.085 & 2.108* & 2.085                     & 8.6    & 133.2 & 4.4                          & 0.057 & 1.359 & 0.088        & 385   & 3600 & 91                 & 4.23      \\ 
                               & (b)    & 2.085 & 2.462* & 2.085                     & 5.5    & 86.8  & 7.0                          & 0.097 & 15.55 & 0.086        & 152   & 3600 & 68                 & 2.24      \\ 
                               & (c)    & 2.085 & 2.098* & 2.085                     & 7.5    & 143.6 & 3.3                          & 0.083 & 0.982 & 0.079        & 493   & 3600 & 105                & 4.70      \\
    \multirow{3}{*}{W-24}      & (a)    & 6.900 & 6.935* & 6.904                     & 22.0   & 76.3  & 11.1                         & 0.086 & 0.586 & 0.086        & 1373  & 3600 & 283                & 4.85      \\ 
                               & (b)    & 6.900 & 7.007* & 6.904                     & 32.7   & 80.0  & 7.8                          & 0.099 & 1.577 & 0.096        & 2856  & 3600 & 1034               & 2.76      \\ 
                               & (c)    & 6.900 & 6.949* & 6.933*                    & 37.4   & 134.4 & 9.2                          & 0.038 & 0.787 & 0.300        & 3401  & 3600 & 3600               & $<$0.94   \\
    \multirow{1}{*}{S-48}      & (b)    & 4.180 & N/A    & 4.180                     & 111.1  & 28.5  & 59.6                         & 0.067 & N/A   & 0.097        & 3600  & 3600 & 2011               & 1.79      \\ \hline
\end{tabular}
\begin{tablenotes}
    \item \noindent
    (a) INC, SOS2: nodal pressures (2x  5 segments), bidirectional average gas flow (10 segments); Z ~5 segments.
    \item \noindent
    (b) INC, SOS2: nodal pressures (2x  7 segments), bidirectional average gas flow (14 segments); Z ~7 segments.
    \item \noindent
    (c) INC, SOS2: nodal pressures (2x 10 segments), bidirectional average gas flow (20 segments); Z ~10 segments.
    \item \noindent
    * Exceeded time limit (3600 seconds).
\end{tablenotes}
\end{threeparttable}
\caption{Model Statistics Summer versus Winter Periods. \label{tab:SvsW-diff}}
\end{table*}
%%%%%%%%%%%%%%%%%%%%%%%%%%%%%%%%%%%%%%%%%%%%%%%%%%%%%%%%%%%%%%%%%%%%%%%%%%%%%
\begin{table*}[!t]
\footnotesize
\centering
\begin{threeparttable}
\begin{tabular}{|c|c c c|c c c|c c c|c c c|c|}
\hline
    \multicolumn{1}{|c|}{Case}         & \multicolumn{3}{c|}{Objective Function}    & \multicolumn{3}{c|}{\# Nodes $[\times 10^3]$} & \multicolumn{3}{c|}{Gap}  & \multicolumn{3}{c|}{CPU Time} & Speed-up  \\ 
    \multicolumn{1}{|c|}{W-24 (a)}     & \multicolumn{3}{c|}{[M\$]}                 & \multicolumn{3}{c|}{[--]}                     & \multicolumn{3}{c|}{[\%]} & \multicolumn{3}{c|}{[s]}      & [--]     \\ \cline{2-14}
    ~                                  & INC    & SOS2    & Z                       & INC    & SOS2 & Z                             & INC   & SOS2   & Z        & INC   & SOS2 & Z              & --       \\ \hline
    \multirow{1}{*}{0.01\%}            & 6.911  & 6.948*  & 6.914                   & 41.7   & 56.8 & 44.9                          & 0.086 & 1.096  & 0.094    & 2574  & 3600 & 1132           & 2.27     \\ 
    \multirow{1}{*}{0.1\% }            & 7.006* & 7.140*  & 6.999                   & 37.1   & 70.1 & 32.2                          & 0.228 & 2.481  & 0.099    & 3600  & 3600 & 672            & $>$ 5.36 \\ 
    \multirow{1}{*}{1\%   }            & 7.937* & 9.302*  & 7.853                   & 51.4   & 93.5 & 31.3                          & 0.925 & 23.591 & 0.099    & 3600  & 3600 & 448            & $>$ 8.04 \\ \hline
\end{tabular}
\begin{tablenotes}
    \item \noindent
    * Exceeded time limit (3600 seconds).
\end{tablenotes}
\end{threeparttable}
\caption{Model Statistics Winter Day -- Sensitivity Losses. \label{tab:Loss}}
\end{table*}
%%%%%%%%%%%%%%%%%%%%%%%%%%%%%%%%%%%%%%%%%%%%%%%%%%%%%%%%%%%%%%%%%%%%%%%%%%%%%
\begin{table*}[!t]
\footnotesize
\centering
\begin{threeparttable}
\begin{tabular}{|c c|c c c|c c c|c c c|c c c|c|}
\hline
    \multicolumn{2}{|c|}{Case}                  & \multicolumn{3}{c|}{Objective Function}   & \multicolumn{3}{c|}{\# Nodes $[\times 10^3]$} & \multicolumn{3}{c|}{Gap}  & \multicolumn{3}{c|}{CPU Time} & Speed-up  \\ 
    \multicolumn{2}{|c|}{}                      & \multicolumn{3}{c|}{[M\$]}                & \multicolumn{3}{c|}{[--]}                     & \multicolumn{3}{c|}{[\%]} & \multicolumn{3}{c|}{[s]}      & [--]     \\ \cline{3-15}
    \multicolumn{2}{|c|}{}                      & INC    & SOS2   & Z                       & INC    & SOS2  & Z                            & INC   & SOS2  & Z         & INC   & SOS2 & Z              & --       \\ \hline
    \multirow{2}{*}{S-24 (a)} & identical       & 2.080  & 2.093* & 2.080                   & 9.7    & 123.4 & 8.5                          & 0.075 & 0.879 & 0.082     & 395   & 3600 & 242            & 1.63     \\ 
                              & very different  & 2.513  & 2.522* & 2.513                   & 4.1    & 153.4 & 5.0                          & 0.030 & 0.448 & 0.002     & 83    & 3600 & 125            & 0.66     \\ \hline                  
    \multirow{2}{*}{W-24 (a)} & identical       & 6.889  & 6.955* & 6.894                   & 13.6   & 81.2  & 39.6                         & 0.093 & 1.009 & 0.087     & 712   & 3600 & 833            & 0.85     \\
                              & very different  & 7.843* & 7.887* & 7.946                   & 113.9  & 94.6  & 103.1                        & 0.195 & 1.181 & 0.097     & 3600  & 3600 & 2018           & $>$ 1.78 \\ \hline               
\end{tabular}
\begin{tablenotes}
    \item \noindent
    * Exceeded time limit (3600 seconds).
\end{tablenotes}
\end{threeparttable}
\caption{Model Statistics -- Sensitivity Gas Cost. \label{tab:GasCost}}
\end{table*}
\end{landscape}

\subsubsection{Operational Results}
\label{sec:OperationalResults}
\begin{figure}
\centering
\includegraphics[scale=1.2]{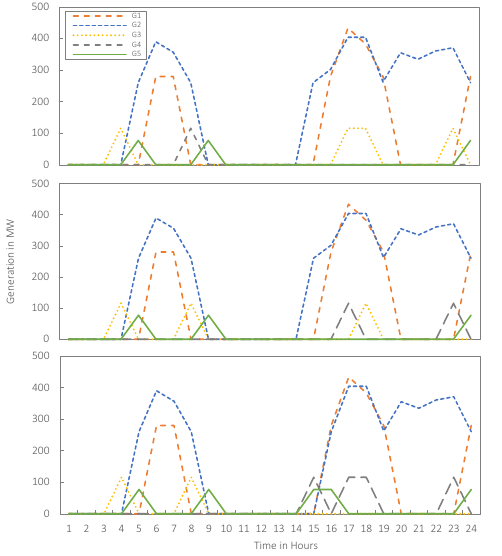}
\caption{Generation of gas-fired units under the INC (top), SOS2 (middle) and Z (bottom) framework for W-24 (a).}
\label{fig:Gas-Gen}
\end{figure}

Table~\ref{tab:Operation} illustrates gas-related operational results for S-24 and W-24 case~(a) and moderately different gas cost\footnote{All cases are solved within the relative MILP gap of 0.1\%. Solution time for the winter day under SOS2 is more than 60 hours.}. For both cases, the total linepack, particularly under INC and SOS2, is significantly different, despite being equivalent formulations. This is due to the fully endogenous pressure decisions, i.e., no fixed slack pressure due to the assumption of downstream pressure regulating equipment, which enables the highest level of linepack flexibility. In general, Z results in lower total linepack and thus lower absolute pressures. This is due to the implied uniform character of the average gas flow over the absolute pressure (see Fig.~\ref{fig:Quality-Lin}).
However, gas production and gas-fired generation are very similar across the three frameworks. Considering INC as the base case, Z yields identical results for the summer day, except for the generation of G3 and G4, which is reversed. For the winter day, differences per gas source and generator are more pronounced, as illustrated by~Table~\ref{tab:Operation} and Fig.~\ref{fig:Gas-Gen}, which shows gas-fired generation over time. However, total gas-fired generation under Z is less than 0.5\% off compared to INC. To evaluate the feasibility of the operational decisions determined under Z, we fix all relevant operating decisions in the power sector, i.e., thermal and renewable generation and power flows, determined under INC and re-run the model under the Z framework. The resulting difference in total system cost is +0.008~M\$ (+0.116\%), which we consider acceptable.\\
As a general takeaway, compared to INC and SOS2, the tightness, i.e., the reduced feasible region due to integrality of binary variables in the polyhedron, and the compactness with respect to the number of variables of the Z framework result in improved computational performance at an acceptable difference of operational decisions and total system cost.

\begin{landscape}
\begin{table*}[!t]
\footnotesize
\centering
\begin{tabular}{|c c| c c c c c c c c c|}
\hline
\multicolumn{2}{|c|}{Case}                    & Total Linepack            & \multicolumn{3}{c}{Gas Production}           & \multicolumn{5}{c|}{Gas-fired Generation}  \\
\multicolumn{2}{|c|}{}                        & [MSm\textsuperscript{3}]  & \multicolumn{3}{c}{[MSm\textsuperscript{3}]} & \multicolumn{5}{c|}{[MWh]}                 \\ \cline{3-11}
\multicolumn{2}{|c|}{}           &                                        & GS1   & GS2    & GS3                         & G1   & G2   & G3   & G4   & G5             \\ \hline
\multirow{3}{*}{S-24 (a)}        & INC       & 51.265                     & 0.000 & 3.419  & 0.837                       & 1120 & 2705 & 348  & 1167 & 308            \\
                                 & SOS2      & 82.390                     & 0.000 & 3.385  & 0.872                       & 1120 & 3096 & 973  & 232  & 308            \\
                                 & Z         & 17.186                     & 0.000 & 3.419  & 0.837                       & 1120 & 2705 & 1167 & 348  & 308            \\ \hline         
\multirow{2}{*}{W-24 (a)}        & INC       & 83.483                     & 0.246 & 11.053 & 3.043                       & 2216 & 4580 & 464  & 116  & 231            \\
                                 & SOS2      & 97.617                     & 0.341 & 10.938 & 3.064                       & 2216 & 4580 & 348  & 232  & 231            \\
                                 & Z         & 56.305                     & 0.395 & 10.349 & 3.605                       & 2216 & 4278 & 232  & 464  & 385            \\ \hline
\end{tabular}
\caption{Operational Results. \label{tab:Operation}}
\end{table*}
\end{landscape}

\section{Conclusions}
\label{sec:Conclusions}
This paper presented the Z piecewise linearization method for the pipeline gas transmission problem with linepack. The proposed Z formulation is specifically designed to be tight, which we showed numerically. The convex hull proof, however, was beyond the scope of this paper. The tightness of Z reduces the solution space to be explored by a MILP solver. In addition, Z is more compact in the number of variables than other piecewise linearizations in the literature, e.g., INC and SOS2. As a result of these characteristics, on average Z can explore more nodes per time during the branch and bound phase, suggesting that the reduced LP relaxations repeatedly solved during this phase are solved faster. This could be due to the reduced number of vertices of the feasible region of Z (see Table~\ref{tab:Polyhedra}), which potentially speeds up the search for the LP optimum of the simplex algorithms.
Our numerical results of an integrated power and gas energy system optimization model illustrate that Z enables finding high-quality solutions and solving the given problem instances 2.57 times faster on average. However, compared to existing methods, Z is a more conservative approximation of gas flows, which results in a moderate difference in the determined operational decisions and the objective function value.
In future research, we plan to further improve the quality of the Z linearization in terms of gas flow physics and extend it to include pipeline transmission expansion planning while maintaining its beneficial characteristics with respect to solution time.

\section*{Acknowledgments}
\label{sec:Acknowledgements}
This work is part of the project iKlimEt (FO999910627), which has received funding in the framework of ”Energieforschung”, a research and technology program of the Klima- und Energiefonds. The work of G. Morales-España was supported by the European Climate, Infrastructure and Environment Executive Agency under the European Union’s HORIZON Research and Innovation Actions under grant agreement No. 101095998. T. Klatzer gratefully acknowledges the Netherlands Organisation for applied scientific research TNO for hosting his research visit and the funding granted by the Rudolf Chaudoire Foundation and the Erasmus+ program.

\appendix
\section{Algorithm for Apriori Parameter Computation}
\label{sec:Appendix}
This section presents a verbal description of the algorithm for the apriori calculation, i.e., before running the ESOM, of parameters $\Tilde{A}_{l,z,\Tilde{z}}$ to $\Tilde{F}_{l,u,v,z}$ including $Z^{Aux}_{l,u,v,z,w}$, $Z^{sgn}_{l,u,v,w}$, $Z^{RHS}_{l,u,v,w}$ and $Z^{Pre}_{l,u,v,z,w}$ relevant for constraints~\eqref{eqn:IEQ4}-\eqref{eqn:IEQ9}. The algorithm is implemented in Python and accessible in a GitHub repository~\cite{GitHubRepo_DataPre_2024}.

The following calculations are done for every pipeline $l$.
\begin{enumerate}
    \item We calculate $\Tilde{A}_{l,z,\Tilde{z}}$, $\Tilde{B}_{l,\Tilde{z},z}$, and $\Tilde{C}_{l,z,\Tilde{z}}$ where $\Tilde{z}$ is an alias of $z\in\mathcal{Z}$: \\
    $\Tilde{A}_{l,z,\Tilde{z}} = min(P_{l,z} F_{l,\Tilde{z}} - F_{l,z} P_{l,\Tilde{z}}, 0) \quad \forall~z\in\mathcal{Z},~z \neq \Tilde{z}$ \\
    $\Tilde{B}_{l,\Tilde{z},z} = max(F_{l,z} P_{l,\Tilde{z}} - P_{l,z} F_{l,\Tilde{z}}, 0) \quad \forall~z\in\{2,\dots,Z\} \leq \Tilde{z}$ \\
    $\Tilde{C}_{l,z,\Tilde{z}} = min(F_{l,z} P_{l,\Tilde{z}} - P_{l,z} F_{l,\Tilde{z}}, 0) \quad \forall~z\in\{2,\dots,Z\} \leq \Tilde{z}$
    \item To determine $Z^{Aux}_{l,u,v,z,w}$, $Z^{sgn}_{l,u,v,w}$, $Z^{RHS}_{l,u,v,w}$ and $Z^{Pre}_{l,u,v,z,w}$, we introduce index $u\in\mathcal{U}=\{1,2,\dots,max(F_{l,z=Z}\!-\!F_{l,z=2}, P_{l,z=Z}\!-\!P_{l,z=2})\}$ with alias $v$.
    \item We then calculate $Z_{l,u,v,z} = u F_{l,z} - v P_{l,z}$
    $\quad \forall~l(m,n)\in\mathcal{L},~u,v\in\mathcal{U},~z\in\mathcal{Z}$.
    \item We iterate through the tuples $(u,v,z)$ of $Z_{l,u,v,z}$ to identify tuples that have at least 2 identical values in $Z_{l,u,v,z}$ and set $Z_{l,u,v,z}=0$ for all tuples where $Z_{l,u,v,z}$ of the current tuple is a multiple (element-wise) of $Z_{l,u,v,z}$ of any previous tuple.
    \item From this reduced $Z_{l,u,v,z}$ we set $\Hat{Z}_{l,u,v,z}=Z_{l,u,v,z}$ \\
    $\forall~z\in\mathcal{Z},~z\neq\Tilde{z},~Z_{l,u,v,z}=Z_{l,u,v,\Tilde{z}}$.
    \item Tuples $(u,v,z)$ of $\Hat{Z}_{l,u,v,z}$ can contain up to $\lfloor \frac{Z-1}{2} \rfloor$ sets with at least 2 identical values for $\Hat{Z}_{l,u,v,z}$. With the index $w\in\mathcal{W} = \{1,2,\dots,\lfloor\frac{Z-1}{2} \rfloor\}$ we partition these tuples such that every set with at least 2 identical values for $\Hat{Z}_{l,u,v,z}$ is mapped as a distinct tuple in $Z^{Aux}_{l,u,v,z,w}$.
    \item In an interim step in the partitioning, we calculate $Z^{sgn}_{l,u,v,w}$ and $Z^{RHS}_{l,u,v,w}$, where $Z^{RHS}_{l,u,v,w} = - abs(Z^{sgn}_{l,u,v,w})$.
    \item For $w=\{2,\dots, \lfloor \frac{Z-1}{2} \rfloor\}$ we assign $Z_{l,u,v,z,w}=Z_{l,u,v,z}$ and filter $Z_{l,u,v,z,w}$ using $Z^{Aux}_{l,u,v,z,w}$ to derive $Z^{Pre}_{l,u,v,z,w}$.
    \item Finally, we calculate $\Tilde{D}_{l,u,v,z}$, $\Tilde{E}_{l,u,v,z}$ and $\Tilde{F}_{l,u,v,z}$: \\    
    $\Tilde{D}_{l,u,v,z} = \begin{cases}
    + u F_{l,z} - v P_{l,z} &\text{if} \ Z^{Pre}_{l,u,v,z,w} \leq Z^{Aux}_{l,u,v,z,w} \
    \text{and} \ Z^{Aux}_{l,u,v,z,w} > 0 \\
    - u F_{l,z} + v P_{l,z} &\text{if} \ Z^{Pre}_{l,u,v,z,w} \geq Z^{Aux}_{l,u,v,z,w} \
    \text{and} \ Z^{Aux}_{l,u,v,z,w} < 0 \\
    {0} &\text{otherwise}  
    \end{cases}$ \\
    $\Tilde{E}_{l,u,v,z} = \begin{cases}
    + Z^{Aux}_{l,u,v,z,w} &\text{if} \ Z^{Pre}_{l,u,v,z,w} > Z^{Aux}_{l,u,v,z,w} \
    \text{and} \ Z^{Aux}_{l,u,v,z,w} > 0 \\
    - Z^{Aux}_{l,u,v,z,w} &\text{if} \ Z^{Pre}_{l,u,v,z,w} < Z^{Aux}_{l,u,v,z,w} \
    \text{and} \ Z^{Aux}_{l,u,v,z,w} < 0 \\
    {0} &\text{otherwise} 
    \end{cases}$ \\
    $\Tilde{F}_{l,u,v,z} = \begin{cases}
     + Z^{Aux}_{l,u,v,z,w} - (u F_{l,z} - v P_{l,z})             \\
       \text{if}  \ Z^{Pre}_{l,u,v,z,w} > Z^{Aux}_{l,u,v,z,w}    \\
       \text{and} \ Z^{Aux}_{l,u,v,z,w} > 0                      \\
     - Z^{Aux}_{l,u,v,z,w} + (u F_{l,z} - v P_{l,z})             \\ 
       \text{if}  \ Z^{Pre}_{l,u,v,z,w} \leq Z^{Aux}_{l,u,v,z,w} \\
       \text{and} \ Z^{Aux}_{l,u,v,z,w} < 0                      \\
      {0} \ \ \text{otherwise} 
    \end{cases}$
\end{enumerate}

\section{Logic for Piecewise Linearization}
\label{sec:Logic}
\begin{figure*}[!t]
\centering
\includegraphics[scale=0.8]{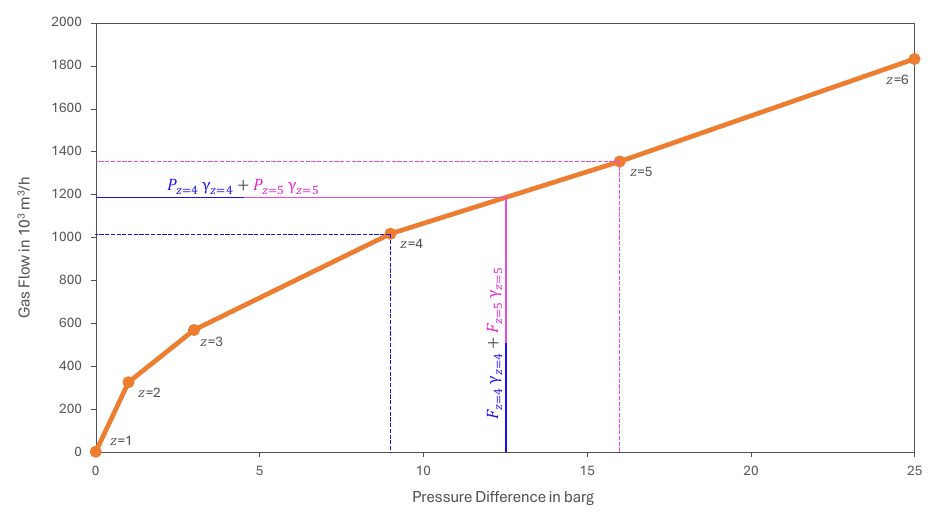}
\caption{Logic of the Z piecewise linearization method.}
\label{fig:LinLogic}
\end{figure*}

The key logic to model the average gas flow and pressure difference is to ensure both are constrained to follow the specified piecewise linear segments representing their relation. To achieve this, the sum of the grid point activation variables $\gamma_{k,l,z}$~\eqref{eqn:LOUP_gamma} must be equal to 1~\eqref{eqn:Sum_gamma} and only adjacent $\gamma_{k,l,z}$ must take values greater than zero. Adjacency is enforced by the binary variable~$\delta_{k,l,z}$~\eqref{eqn:Def_delta} via~\eqref{eqn:SOS2Logic}, where the sum of all~$\delta_{k,l,z}$ is also equal to~1~\eqref{eqn:Sum_delta}. Fig.~\ref{fig:LinLogic} illustrates this logic for a case where $\gamma_{k,l,z=4}=\gamma_{k,l,z=5}=0.5$. Accordingly,~\eqref{eqn:SOS2Logic} and~\eqref{eqn:Sum_delta} force~$\delta_{k,l,z=4}=1$ and all~$\delta_{k,l,z\neq4}=0$. Note that for the set of tight constraints in Section~\ref{sec:TightConstraints},~$\delta_{k,l,z}$ always takes binary values, even when defined as a continuous variable.

\newpage
\bibliography{main_unmarked}

\end{document}